\documentclass{appolb}
\usepackage{epsfig}
\usepackage{mathrsfs}
\usepackage{cite}
\usepackage{bm,amssymb}
\usepackage{amsmath}
\usepackage{txfonts}
\usepackage{booktabs}
\usepackage{cases}
\def\pbar{\overline{\psi}}

\def\Gs2{\Gamma^{(2)}_{k,\sigma}}
\def\Gp2{\Gamma^{(2)}_{k,\pi}}

\def\ip0{\mathrm{i}p_{0}}
\def\p02{p_{0}^{2}}

\def\q2{\vec{q}^2}

\newcommand{\MeV}{\mathrm{MeV}}
\newcommand{\pab}{
\ifmmode p
\else $p$
\fi
}
\newcommand{\pfour}{
\ifmmode P
\else $P$
\fi
}
\newcommand{\qfour}{
\ifmmode Q
\else $Q$
\fi
}

\begin{document}
\title{Novel picture of the soft modes at the QCD critical point based on the FRG method}
\author{
Takeru Yokota$^{1}$, Teiji Kunihiro$^{1}$ and Kenji Morita$^{2}$
\address{$^1$Department of Physics, Faculty of Science, Kyoto University, Kyoto, Japan\\
 $^2$Yukawa Institute for Theoretical Physics, Kyoto University, Kyoto, Japan}
}
\maketitle
\begin{abstract}
We investigate the soft mode at the QCD critical point (CP) 
on the basis of the functional renormalization group. 
We calculate the spectral functions in the meson channels in the two-flavor quark--meson model.
Our result shows that the energy of the peak position of the particle--hole mode in the sigma
%\sout{mesonic} 
channel becomes vanishingly small as the system approaches the QCD CP, which is a manifestation of the softening of the phonon mode.
We also extract the dispersion curves of the mesonic and the phonon mode, a hydrodynamic mode
which leads to a finding that the dispersion curve of the sigma-mesonic mode crosses the light-cone into the space-like momentum region, and then eventually merges into the phonon mode as the system approaches further close to the CP. This may suggest that the sigma-mesonic mode forms the soft mode together with the hydrodynamic mode
 at the CP.
%\sout{ in contrast to the previous researches.}
\end{abstract}
  
\section{Introduction}
One of the expected structure of the QCD phase diagram is the existence of the first-order phase boundary between the hadronic phase and the quark--gluon plasma phase \cite{Fukushima:2010bq}.
In particular, the phase transition becomes second order at the end point of the phase boundary, which is referred to the QCD critical point (CP).

A system near the CP shows large fluctuations of and correlations between various quantities and thus a method beyond the
mean-field theory is desirable for describing the physical properties near the CP. 
The functional renormalization group (FRG) \cite{Wetterich:1992yh,Wegner:1972ih,Wilson:1974,Polchinski:1983gv}
is a nonperturbative method for the field theory and is expected to be a method to reveal the nature of the system more accurately than the mean-field theory. 
It has been found to be useful in the description of chiral phase transition in QCD via effective chiral models
\cite{Jungnickel:1995fp,Braun:2003ii,Schaefer:2005,Schaefer:2006ds,
Stokic:2010, Nakano:2010, Aoki:2014}.

There exist specific collective modes which are coupled to the fluctuations of the order parameter and become gapless and a long-life at a CP. Such a mode is called the {\em soft mode}.
As for the QCD CP, the nature of the soft modes is nontrivial due to the presence of 
current quark mass together with the violation of charge conjugation symmetry
owing to the finite baryon chemical potential,
%\sout{. In some previous researches}, 
and the soft mode is
considered to be the particle--hole mode corresponding to the density (and energy) 
fluctuations {\cite{Fujii:2004jt,Son:2004iv}}. 

%\sout{Our purpose is to}
{We shall report on our recent work \cite{Yokota:2016tip}, where we have} investigate{d} the nature of
low-energy modes at the QCD CP in the framework of FRG.
We calculate the spectral functions in the sigma 
and pion channels in the two-flavor quark meson model.
Our results confirm the softening of the particle--hole mode 
in the sigma channel near the QCD CP. 
In addition, we find that the low-momentum dispersion
relation of the sigma-mesonic mode penetrates into the space-like momentum region and the mode 
merges into the bump of the particle--hole mode. 
%\sout{This talk is mainly based on our recent work.}

\section{Method}

The FRG is based on the philosophy of the Wilsonian renormalization group 
\cite{Wetterich:1992yh,Wegner:1972ih,Wilson:1974,Polchinski:1983gv}.
In this method, the effective average action (EAA) $\Gamma_{k}$ is 
introduced such that it becomes bare action at a large UV scale $k=\Lambda$ 
and the effective action at $k\rightarrow 0$, where $k$ represents the scale of the renormalization flow.
The flow of EAA for $k$ is 
%\sout{known to be represented} 
{described by} a functional differential equation known 
as the Wetterich equation \cite{Wetterich:1992yh}. 
In principle, the effective action $\Gamma_{k=0}$ is calculated by solving 
the Wetterich equation with the bare action $\Gamma_{\Lambda}$ {being the initial value}.

The spectral functions in the sigma and pion channel 
$\rho_{\sigma, \pi}(\omega, p)$ ($p=|\vec{p}|$)
are obtained in terms of the imaginary parts of the retarded  two-point functions.
The expansion of the Wetterich equation by fields gives infinite series of coupled differential equations for the vertex functions. 
%\sout{Especially} 
{In particular,} the second
%\sout{ary} 
order of the expansion gives 
the flow equation for two-point functions.
%\sout{ because the inverses of the second derivatives of the effective action give the two-point functions.} 
By introducing some approximation 
%{that breaks the coupling of}
{truncating} the infinite series of differential equations, 
the two-point functions can be calculated from the flow equations. 
One of the ways is evaluating the terms in the flow equation using a truncated EAA to solve the flow equation.
Such an approximation scheme has been developed in the cases of the $\mathrm{O}(4)$ model \cite{Kamikado:2013sia} and the quark--meson model \cite{Tripolt:2013jra,Tripolt:2014wra}.

We employ the two-flavor quark--meson model as the low-energy effective model of QCD. This is a chiral effective model consisting of the quark field $\psi$ and mesonic fields $\sigma$ and $\vec{\pi}$.
In our calculation, we employ a truncated EAA for which the local potential approximation is applied in the meson part \cite{Schaefer:2006ds}. With this truncated EAA, we evaluate the terms in the flow equations for the sigma and pion two-point functions.

We employ the imaginary time formalism to analyze the finite-temperature system.
In the formalism, analytic continuation from Matsubara frequency to real frequency is needed to get retarded two-point functions, which often 
%\sout{becomes the difficulty of analyses} 
involves some intricate procedures. In our case, it is known that such a difficulty can be evaded at the level of the flow equations analytically {\cite{Tripolt:2013jra,Tripolt:2014wra, Floerchinger:2011sc}}.

The concrete forms of our flow equations and details for our numerical procedure is presented 
in Ref.~\cite{Yokota:2016tip}.

\section{Result}
We first determine the location of the CP.
Our truncated {EAA} contains the potential term and it provides
static properties of the system \cite{Yokota:2016tip}. From the result of the chiral condensate and the square of the sigma screening mass, i.e. the inverse of the chiral susceptibility, we estimate the critical temperature $T_{c}$ and 
%\sout{critical} 
chemical potential 
$\mu_{c}$ as $(T_{c}, \mu_{c})=(5.1\pm 0.1\,\MeV, 286.6 \pm 0.2\,\MeV)$. In the following discussion,
we regard $T_{c}$ and $\mu_c$ as $5.1\,\MeV$ and
 $286.686\,\MeV$, respectively.

\begin{figure}[!t]
\centering
\epsfig{file=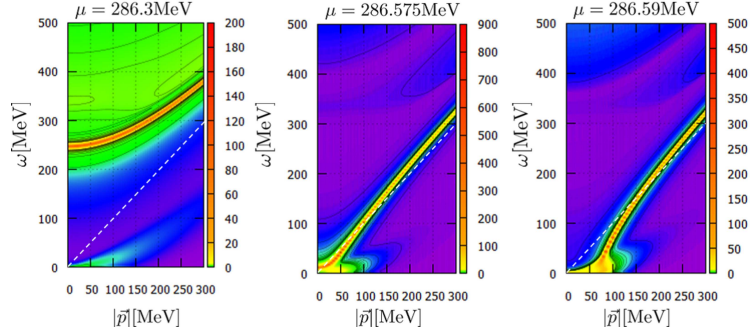,width=0.78\columnwidth}
\caption{Contour maps of $\rho_{\sigma}$
at $T=T_{c}$ and $\mu=286.3\MeV$, $286.575\MeV$ and
$286.59\MeV$. (Taken from \cite{Yokota:2016tip})}
\label{Spect3D}
\end{figure}

%\sout{We show} 
{Now we shall present} the result of the spectral function in the sigma channel $\rho_{\sigma}$ near the QCD CP.
The calculation is performed by increasing the chemical potential toward $\mu_{c}$ 
along a constant temperature line $T=T_{c}$.
Figure~\ref{Spect3D} is the contour map of $\rho_{\sigma}$.
At $\mu=286.3\MeV$, the peaks of the sigma-meosnic mode ($\sigma$ mode) and the particle--hole ({\em p--h}) bump can be seen in the time-like momentum region $\omega>p$ 
%\sout{as well as the particle-hole bump in the} 
and space-like momentum region $\omega<p${, respectively}.
One can also see the dispersion relations of these modes from the peak positions.
As the chemical potential 
%\sout{increases}
{approaches the critical value from below}, the dispersion relation
of the 
%\sout{sigma-mesonic}
$\sigma$ mode shifts downward and it touches the light cone
near $\mu=286.575\MeV$. At $\mu=286.59\MeV$, 
%\sout{in  low-momentum region}
the $\sigma$ mode with small momentum
clearly penetrates into space-like momentum region and
merges to the {\em p--h}
%\sout{particle--hole}
bump, which has a flat dispersion relation in the small momentum region.

\begin{figure}[!t]
\centering
\epsfig{file=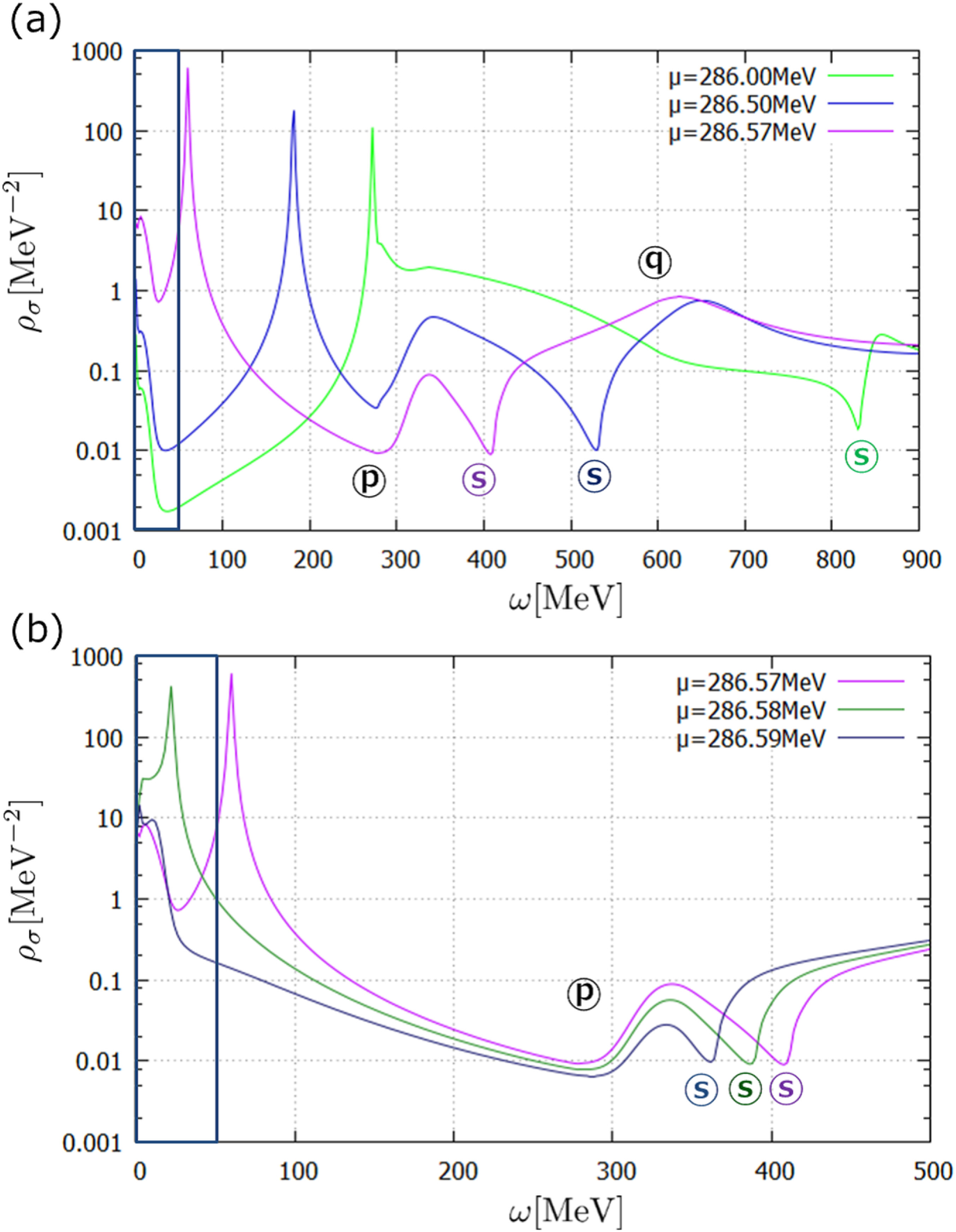,width=0.57\columnwidth}
\caption{The result of $\rho_{\sigma}$ near the QCD CP at $T=T_{c}$ in (a) $286.00\,\MeV \leq \mu \leq 286.57\,\MeV$ and in (b) $286.57\,\MeV \leq \mu \leq 286.59\,\MeV$.
$p$ is set to $50\,\MeV$.
\textcircled{s} represents the $2\sigma$ decay thresholds for each chemical potential. (Taken from  \cite{Yokota:2016tip})}
\label{SpectP50_2}
\end{figure}

Next we show the strength of peaks and bumps of $\rho_{\sigma}$ when $p$ is set to $50 \MeV$. 
The results when $\mu=286.00\MeV$, 
$286.50\MeV$ and $286.57\MeV$
%\sout{$\mu=286.50\MeV$ and $\mu=286.57\MeV$}
are shown in Fig.~\ref{SpectP50_2} (a).
One can see the 
%\sout{sigma-mesonic peak as well as} 
bumps corresponding to $2\sigma$ and $2\pi$ decay in the time-like momentum region 
{as well as the sigma-mesonic peak}.
The peak 
%\sout{position}
of the $\sigma$
%\sout{sigma-mesonic} 
mode shifts to the lower energy as the system approaches the CP.
The position of the $2\sigma$ threshold also goes {down} to a lower energy region
while those of the $2\pi$ and $\pbar \psi$ thresholds hardly
change. The bump in the space-like momentum region is drastically enhanced as the system is close to the CP.
This behavior can be interpreted as the softening of the {\em p--h} 
%\sout{particle--hole} 
mode.
In Fig.~\ref{SpectP50_2}(b), we show the results at chemical potentials much
closer to the CP. Because of numerical instability 
in $286.60\MeV \leq \mu \lesssim 360\MeV$,
we choose $\mu =286.58\MeV$ and
$286.59\MeV$. 
For comparison, the result at $\mu=286.57\MeV$ is also shown.
These results are quite different from those in 
$\mu \leq 286.57\MeV$. In $\mu >286.57\MeV$,
%\sout{the peak of}
the $\sigma$ 
%\sout{sigma-mesonic} 
mode penetrates into the space-like momentum region and then merges into the {\em p--h} 
%\sout{particle--hole} 
mode.
Our results indicate that the $\sigma$
%\sout{sigma-mesonic} 
mode as well as the {\em p--h} 
%\sout{particle--hole} 
mode can become soft at the CP.

%\sout{One of the} 
{A} possible trigger of this phenomenon 
%\sout{is} 
{may be identified with} the  level repulsion between the $\sigma$ 
%\sout{sigma-mesonic} 
mode and other modes.
In particular, the two-sigma ($\sigma\sigma$) 
mode is considered to play an important role since the threshold of 
the $\sigma\sigma$ 
%\sout{two-sigma} 
mode shifts downward as the system approaches the CP.
To see the effect of the repulsion between {the} $\sigma$ and $\sigma\sigma$ mode, we calculate $\rho_{\sigma}$ with changing the strength of the sigma three-point vertex in the flow equation.
Our result 
%\sout{suggests} 
{shows} that the interaction between the two modes 
strongly affects the level of the $\sigma$
%\sout{sigma-mesonic} 
mode.

We also calculate the spectral function in {the} pion channel $\rho_{\pi}$.
No critical behavior of modes is observed in $\rho_{\pi}$ {in contrast to the $\sigma$ channel}.

\section{Summary}

We have calculated the spectral functions in the sigma and pion channels with the functional renormalization group 
%\sout{method}
to analyze the soft mode at the QCD CP. 
We have employed the two-flavor quark--meson model, and our calculation 
is based on the local potential approximation. The bump of the particle--hole mode 
in the spectral function in the sigma
%\sout{meson} 
channel is enhanced 
as the system approaches the CP, which corresponds to the softening of {the} phonon mode. 
We have also found that the dispersion relation of the sigma-mesonic mode penetrates 
into the space-like momentum region and merges into the particle--hole mode 
as the system further approaches 
%\sout{to} 
the CP, which suggests the softening of the sigma-mesonic mode.

\end{document}